\documentclass[conference]{IEEEtran}
\IEEEoverridecommandlockouts

\PassOptionsToPackage{hyphens}{url}

\usepackage{amsmath,amssymb,amsfonts,amsthm}
\usepackage{bm}
\usepackage{textcomp}

\usepackage{graphicx}
\usepackage[caption=false,font=normalsize,labelfont=sf,textfont=sf]{subfig}
\usepackage{stfloats}
\usepackage{array,booktabs,tabularx}

\usepackage{algorithm}
\usepackage{algorithmic}
\usepackage{verbatim}

\usepackage{cite}
\usepackage{xcolor}
\usepackage{url,xurl}
\usepackage{microtype}
\usepackage{enumitem}
\usepackage{etoolbox}

\setlist{noitemsep,topsep=0pt,parsep=0pt,partopsep=0pt}

\newcommand{\compactdisplay}{%
  \setlength{\abovedisplayskip}{2pt plus 0pt minus 1pt}%
  \setlength{\belowdisplayskip}{2pt plus 0pt minus 1pt}%
  \setlength{\abovedisplayshortskip}{1pt plus 0pt minus 1pt}%
  \setlength{\belowdisplayshortskip}{1pt plus 0pt minus 1pt}%
  \setlength{\jot}{0pt}%
}

\AtBeginEnvironment{align}{\compactdisplay}
\AtBeginEnvironment{align*}{\compactdisplay}
\AtBeginEnvironment{equation}{\compactdisplay}
\AtBeginEnvironment{equation*}{\compactdisplay}
\AtBeginEnvironment{gather}{\compactdisplay}
\AtBeginEnvironment{gather*}{\compactdisplay}
\AtBeginEnvironment{multline}{\compactdisplay}
\AtBeginEnvironment{multline*}{\compactdisplay}

\newtheorem{theorem}{Theorem}

\newtheorem{proposition}{Proposition}

\newcommand{\LMP}{\mathsf{LMP}}
\newcommand{\LMPUm}{\mathsf{LMP+U}}
\newcommand{\LMPMWP}{\mathsf{LMP+MWP}}

\newcommand{\MWP}{\mathsf{MWP}}
\newcommand{\MS}{\mathsf{MS}}

\renewcommand{\baselinestretch}{0.99}
\begin{document}
\raggedbottom
\title{Eliminating Make-Whole Uplifts \\with a Minimum-Cost Uniform Price Adder
}
\author{
Valentina Norambuena-Guzman,~\IEEEmembership{Student Member,~IEEE},
\quad
 Cong Chen,~\IEEEmembership{Member,~IEEE},
  \\
 \qquad
 Lang Tong,~\IEEEmembership{Fellow,~IEEE},
 and Timothy D. Mount,~\IEEEmembership{Life Member,~IEEE}

\thanks{\scriptsize 
Valentina Norambuena and Lang Tong (\{van27,lt35\}@cornell.edu) are with the School of Electrical and Computer Engineering, Cornell University, Ithaca, NY, USA.
Cong Chen (cong.chen@dartmouth.edu) is with the Thayer School of Engineering, Dartmouth College, Hanover, NH, USA. 
Timothy D. Mount (tdm2@cornell.edu) is with the Dyson School of Applied Economics and Management, Cornell University, Ithaca, NY, USA. 
This work is supported in part by the National Science Foundation under Awards 2603293 and 2412776, and the Power Systems Engineering Research Center (PSERC) Research Project M-48.}}



\maketitle

\begin{abstract}
Co-optimizing energy and flexible ramping reserves may dispatch generators at locational marginal prices (LMPs) below their submitted energy offer prices, necessitating out-of-market make-whole payments. Such payments create discriminatory settlements among generators and may incentivize strategic bidding for higher profits. This paper proposes LMP+U—an in-market uniform price adder to LMP for transmission-constrained networks—which minimizes demand payments subject to zero out-of-market make-whole uplift, preserves LMP-based congestion charges, and provides locally truthful bidding incentives for price-taking generators. We also identify conditions under which a price-taking generator can increase its profit by inflating its energy offer under conventional LMP with make-whole uplift. Numerical simulations illustrate the properties of LMP+U and its advantages over conventional LMP in congested networks.
\end{abstract} 

\section{Introduction}
The flexible ramping product (FRP) is a market mechanism that co-optimizes energy dispatch and ramping reserves for markets with substantial and uncertain ramping needs.  The California Independent System Operator (CAISO) incorporates FRP into multi-interval look-ahead optimization \cite{CAISO2023FRP}, whereas the Midcontinent Independent System Operator (MISO) uses single-interval real-time dispatch with ramp capability requirements covering forecast net-demand changes and uncertainty beyond the dispatch interval \cite{MISO2016Ramp,MISO2025FRP}.  In both cases, forecasts within the look-ahead window are involved in market clearing optimization, which sets the real-time energy and ramp-reserve prices.

Because rolling-window dispatch optimizes schedules over a look-ahead horizon but settles only the binding interval, intertemporal ramping constraints can require generators to produce at locational marginal prices (LMPs) below their submitted energy offers, resulting in out-of-merit dispatch. Revenues anticipated in later intervals are not guaranteed, since future schedules and prices remain advisory. Ramping reserve requirements, which may impose positive lower bounds on generator outputs, can further contribute to such dispatch. To compensate generators for energy bid-cost shortfalls, system operators employ bid-cost recovery mechanisms, typically in the form of out-of-market make-whole payments (MWPs). The costs of these settlements \cite{CAISORpt:22} can be substantial and are generally recovered through charges to market participants.

Resource-specific out-of-market uplifts raise several concerns regarding key objectives of real-time electricity price formation articulated by FERC. Specifically, effective price formation should (a) promote transparency by ensuring that real-time prices reflect the marginal cost of serving load and the operational constraints of the system \cite{FERC15}, whereas resource-specific out-of-market MWPs create discriminatory effective energy prices that are not reflected in publicly observable market prices; (b) provide appropriate incentives for market participants to follow dispatch instructions \cite{FERC15}, whereas LMP alone may fail to compensate generators adequately for following dispatch instructions, necessitating out-of-market MWPs; and (c) promote competitive market outcomes by encouraging generators to submit energy offers reflecting their marginal costs \cite{FERC14}, whereas out-of-market MWPs may incentivize price-taking generators to inflate their energy offers to increase profits. Such incentives were previously identified for lost opportunity cost (LOC) uplifts in \cite{Guo21TPS} and are investigated for energy make-whole payments in Sec.~\ref{sec:incentives} of this paper.



This paper develops an in-market solution for FRP that eliminates out-of-market settlement with the minimum-cost uniform (thus non-discriminative) pricing at each bus. Eliminating out-of-market uplift, however, does not by itself guarantee truthful bidding incentives. We therefore investigate whether LMP+U provides locally truthful bidding incentives for price-taking generators.

\subsection{Related Work}
The issue of under-compensation under LMP was recognized as early as 2017 \cite{Schiro2017flexibility}, and various uniform pricing mechanisms that minimize but do not eliminate out-of-market settlements have been proposed; see \cite{Hua2019,Guo21TPS} and references therein.  As shown in \cite{Guo21TPS}, uniform nodal pricing cannot generally eliminate lost opportunity costs associated with multi-interval dispatch.  The existence of uniform pricing schemes that remove make-whole uplifts is obvious, although finding one that minimizes demand payments while preserving LMP-based congestion charges is nontrivial. In \cite{Chen25PESGM,Chen26arxiv},
Maximum Dispatch Cost Pricing (MDCP) is proposed for a single-bus system, which eliminates out-of-market MWPs and is shown to provide local truthful-bidding incentives for price-taking generators.  In particular, it is shown that a price-taking generator cannot increase its profit by deviating from the true marginal cost of generation by a limited amount. Generalizing MDCP to multi-bus systems, however, is nontrivial because the notion of ``maximum dispatched cost generator'' is not well defined for multi-bus systems. The work most relevant to this paper is \cite{PESIM}, where LMP+U is proposed to generalize MDCP to multi-bus networks while retaining most desirable properties of MDCP.  In addition, LMP+U has the special property that it preserves the network congestion prices so that, in addressing LMP's weakness in handling inter-temporal constraints, LMP+U retains the spatial property of LMP so that the nodal price difference under LMP+U reflects the marginal cost of transmission constraints and can continue to serve as the basis for settling financial transmission rights.

This paper builds upon the work of \cite{PESIM}, extending its setting to FRP markets along with several new results, as highlighted next.

\subsection{Summary of Contributions}
The main contribution of this work is threefold. First, we extend the LMP+U pricing framework to the energy market with FRP, establishing that LMP with a uniform price adder has the lowest demand cost among all uniform pricing mechanisms that eliminate out-of-market MWP and preserve LMP's congestion price and merchandising surplus.  For a congestion-less network, LMP+U reduces to MDCP \cite{Chen25PESGM}.  When all ramping constraints are non-binding, the uniform adder is zero, and LMP+U reduces to LMP.

Second we address the issue of truthful bidding incentives, showing that a price-taking generator under LMP+U has no incentive to have small deviation from bidding truthfully with its marginal generation cost, a property also shared by LMP\footnote{As in the case of LMP, bidding strategically with a large deviation from a price-taker's marginal cost may increase generator profits. Hence the truthful bidding property is local rather than global for LMP and LMP+U.}. In addition, we identify conditions under which raising an offer of a price-taking generator, while not affecting LMP, could increase generator's net revenue.

Finally, simulations over 100 demand trajectories on the IEEE RTS-24 system examine how FRP procurement, ramping capability, and transmission congestion influence generator net revenues and consumer payments. FRP procurement increases the frequency of energy bid-cost shortfalls, while its net-revenue effects differ across generator groups. Increasing ramping capability can reduce a group's net revenue, and LMP+U under congestion has mixed effects in generators net revenue, preserving LMP locational signals.  Total consumer payments under LMP+U exceed those under LMP+MWP in the cases studied.
\begin{table}[!t]
\caption{Major Symbols and Notation}
\label{tab:major_symbols}
\centering
\footnotesize
\renewcommand{\arraystretch}{0.9}
\setlength{\tabcolsep}{3pt}
\begin{tabularx}{\columnwidth}{@{}l>{\raggedright\arraybackslash}X@{}}
\hline
\textbf{Symbol} & \textbf{Description}\\
\hline
$\mathcal{H}_t,W$
& Look-ahead window and its length in intervals.\\
$[M],\mathcal{N}_m$
& Set of buses and set of generators at bus $m$.\\
$\mathcal{A}_t$
& Set of generators producing energy in interval $t$.\\
$g_{i\tau}^m,\hat d_\tau^m,D_t$
& Scheduled generation, forecast net demand at bus $m$,
and total net demand in interval $t$.\\
$c_{i\tau}^m,c_{i\tau}^{m,\dagger}$
& Submitted energy offer price and true marginal cost.\\
$\boldsymbol\pi_t^{\LMP},\boldsymbol\pi_t^{\LMPUm}$
& LMP and LMP+U energy price vectors.\\
$\Delta\pi_t^*$
& Optimal price adder applied to all nodal LMPs.\\
$\Pi_{it}^m,\mathcal{R}_{it}^m$
& Generator net revenue and FRP payment under the pricing
scheme considered.\\
$\MWP_{it}^m,\MS_t$
& Energy-only MWP and energy merchandising surplus.\\
$\delta,\alpha_\ell,\rho_{it}^m$
& Energy offer price increase, and LMP and FRP-payment
sensitivities to that increase.\\
\hline
\end{tabularx}

\smallskip
\begin{minipage}{\columnwidth}
\footnotesize
Index $i$ denotes a generator; $m$ and $\ell$ denote buses.
Index $t$ denotes the binding interval, and $\tau$ indexes
intervals in $\mathcal{H}_t$. Superscripts $R$ and $*$
denote implemented decisions and optimal values, respectively.
The interval duration is included in cost, payment,
and net revenue calculations but is not written explicitly.
\end{minipage}
\end{table}
\section{Economic Dispatch and Settlement}
\label{sec:model}
\subsection{Rolling-Window Economic Dispatch}
At each interval $t$, the operator schedules energy and flexible ramping capability over the look-ahead window $\mathcal H_t:=\{t,\ldots,t+W-1\}$. Only decisions for the binding interval $t$ are implemented and settled; those for subsequent intervals are advisory. The dispatch is re-solved each interval, with ramping constraints initialized from the preceding interval's implemented generation.

Table~\ref{tab:major_symbols} summarizes the main notation. For each $\tau\in\mathcal H_t$, the dispatch takes forecast net demand $\hat d_\tau^m$ and submitted energy offer prices $c_{i\tau}^m$ as inputs. The operator schedules generation $g_{i\tau}^m$ and awards upward and downward FRP quantities $\bar r_{i\tau}^m$ and $\underline r_{i\tau}^m$, respectively.

FRP reserves generation capability to accommodate forecasted net demand changes and forecast errors. Case $k=0$ represents the scheduled dispatch at forecast net demand, while cases $k=U,D$ represent the higher and lower net-demand levels covered by the FRP requirements. In these two cases, generation increases $x_{i\tau}^{U,m}$ and decreases $x_{i\tau}^{D,m}$ relative to the scheduled dispatch must balance the corresponding net-demand changes and remain within the awarded FRP quantities. For each case, $\boldsymbol z_\tau^k$ denotes the vector of nodal net injections, defined as generation minus net demand. The constraints $S\boldsymbol z_\tau^k\leq\bar{\boldsymbol f}$ enforce transmission limits for the scheduled dispatch and both FRP cases, ensuring that the reserved capability can accommodate the specified net-demand changes without overloading transmission lines. Building on \cite{Chen26arxiv,PESIM}, the simplified dispatch problem is:
\begingroup
\setlength{\jot}{1.5pt}
\begin{subequations}
\label{eq:dispatch}
\begin{align}
\mathsf{ED}_t:\quad
&\min_{\substack{g,\bar r,\underline r,\\x^U,x^D}}
\sum_{\tau\in\mathcal H_t}
\sum_{m\in[M]}
\sum_{i\in\mathcal N_m}
c_{i\tau}^m g_{i\tau}^m
\label{eq:objective}\\
\text{s.t.}\quad
&\sum_{m\in[M]}\hat d_\tau^m
-\sum_{m\in[M]}\sum_{i\in\mathcal N_m}g_{i\tau}^m=0,
\quad(\lambda_\tau)
\label{eq:balance}\\
&S\boldsymbol z_\tau^k\leq\bar{\boldsymbol f},
\quad k\in\{0,U,D\},\quad(\boldsymbol\phi_\tau^k)
\label{eq:network}\\
&\text{capacity, ramping, and FRP constraints.}
\nonumber
\end{align}
\end{subequations}
\endgroup
All constraints apply to each $\tau\in\mathcal H_t$. Here, $\lambda_\tau$ is the energy-balance multiplier, $S$ is the shift-factor matrix accounting for both flow directions, $\bar{\boldsymbol f}$ is the vector of corresponding line limits, and $\boldsymbol\phi_\tau^k$ is the vector of transmission-constraint multipliers for case $k$. 
\subsection{Energy Prices, FRP Payments, and MWPs}
Energy and FRP are settled for the binding interval $t$, with $\hat d_t^m$ taken as the settled net demand at bus $m$. With FRP requirements and their allocation across buses fixed, the nodal LMP vector is obtained from the dual solution of \eqref{eq:dispatch}:
\begin{equation}
\boldsymbol\pi_t^{\LMP}
=\lambda_t^*\boldsymbol 1
-S^\top\bigl(\boldsymbol\phi_t^{0*}
+\boldsymbol\phi_t^{U*}
+\boldsymbol\phi_t^{D*}\bigr),
\label{eq:lmp}
\end{equation}
where superscript $*$ denotes optimal multipliers of \eqref{eq:dispatch}. The LMPs combine the energy-balance component with congestion components from the scheduled dispatch and both FRP cases. The latter reflect transmission constraints on FRP deliverability. 

Generator $i$ at bus $m$ receives an FRP payment $\mathcal R_{it}^m$, equal to the sum of its upward and downward FRP awards multiplied by their respective nodal FRP prices. These prices reflect ramping procurement requirements and transmission deliverability constraints \cite{CAISO2023FRPDeliverability}. 


Ramping and FRP constraints can require generation at an LMP below the submitted energy offer price \cite{Chen26arxiv,PESIM}. For a nodal energy price $\pi_{mt}$, the energy-only MWP covers any shortfall between submitted energy costs and energy revenue:
\begin{equation}
\MWP_{it}^m(\pi_{mt})
=\max\{(c_{it}^m-\pi_{mt})g_{it}^{m,R},\,0\},
\label{eq:mwp}
\end{equation}
where $g_{it}^{m,R}$ is the binding-interval generation implemented from the optimal solution of \eqref{eq:dispatch}. The LMP+MWP benchmark evaluates \eqref{eq:mwp} at $\pi_{mt}=\pi_{mt}^{\LMP}$.

Let $c_{it}^{m,\dagger}$ denote the generator's true marginal production cost. Its net revenue under LMP+MWP equals energy revenue plus the MWP and FRP payments, less production costs:
\begin{equation}
\Pi_{it}^{m,\LMPMWP}
=\bigl(\max\{c_{it}^m,\pi_{mt}^{\LMP}\}
-c_{it}^{m,\dagger}\bigr)g_{it}^{m,R}
+\mathcal R_{it}^m.
\label{eq:lmp_profit}
\end{equation}
For any nodal energy price vector $\boldsymbol\pi_t$, energy merchandising surplus is the difference between total energy payments collected from net demand and those paid to generators:
\begin{equation}
\MS_t(\boldsymbol\pi_t)
=\sum\nolimits_m\pi_{mt}\hat d_t^m
-\sum\nolimits_{m,i}\pi_{mt}g_{it}^{m,R}.
\label{eq:merchandising_surplus}
\end{equation}
This surplus provides funds for market settlement obligations. A pricing scheme that ensures energy bid-cost recovery while preserving this surplus maintains the funds available for those obligations \cite{PESIM}.

\section{Optimal Uniform Pricing without MWP}
\label{sec:lmpu}
Building on~\cite{PESIM}, LMP+U follows the uniform-price auction principle that accepted energy offers are covered by a common price at each bus. We minimize binding-interval energy payments by demand subject to energy bid-cost recovery and preservation of nodal LMP differences, retaining congestion charges for transfers between buses. Preserving these differences requires a common adjustment to every nodal LMP, which we call the uniform price adder.

\subsection{Pricing Problem and Optimality of LMP+U}
Consider the binding-interval dispatch from Sec.~\ref{sec:model}. Generation, FRP awards and payments, and the LMP vector are fixed during the pricing step. Let $\mathcal A_t$ denote the generators producing energy in interval $t$, and let $D_t:=\sum_m\hat d_t^m$ denote total net demand. Following \cite{PESIM}, we choose nodal energy prices by solving
\begin{subequations}
\label{eq:pricing_problem}
\begin{align}
\min_{\boldsymbol\pi_t}\quad
&\sum_m\pi_{mt}\hat d_t^m\\
\text{s.t.}\quad
&\pi_{mt}\geq c_{it}^m,
\quad (i,m)\in\mathcal A_t,\\
&\pi_{mt}-\pi_{nt}
=\pi_{mt}^{\LMP}-\pi_{nt}^{\LMP},
\quad m,n\in[M].
\end{align}
\end{subequations}
\begin{theorem}[Optimal uniform energy pricing]
\label{thm:optimal_lmpu}
If $D_t>0$, problem \eqref{eq:pricing_problem} has the unique solution
\begin{align}
\Delta\pi_t^*
&=\max_{(i,m)\in\mathcal A_t}
\{c_{it}^m-\pi_{mt}^{\LMP}\},
\label{eq:adjustment}\\
\boldsymbol\pi_t^{\LMPUm}
&=\boldsymbol\pi_t^{\LMP}+\Delta\pi_t^*\boldsymbol1.
\label{eq:lmpu}
\end{align}
\end{theorem}
The proof follows the argument in \cite{PESIM}. The uniform price adder $\Delta\pi_t^*$ is the common offset applied to every nodal LMP in \eqref{eq:lmpu}. A generator attaining the maximum in \eqref{eq:adjustment} is an adder setter, ties are possible. The adder is negative when every dispatched generator's LMP exceeds its submitted energy offer price.

\subsection{Properties}
LMP+U has the following properties:
\begin{enumerate}
\item Each dispatched generator's energy revenue covers its submitted energy offer cost, eliminating  MWPs.
\item LMP+U preserves nodal price differences and congestion charges. Under energy balance, the adder changes total energy payments collected from net demand and paid to generators by the same amount. Hence,
\[
\MS_t(\boldsymbol\pi_t^{\LMPUm})
=\MS_t(\boldsymbol\pi_t^{\LMP}).
\]
\item When LMPs are equal at all buses, LMP+U becomes MDCP: the energy price at every bus equals the highest submitted energy offer price among dispatched generators. This holds, in particular, when all transmission constraints for energy and upward and downward FRP have zero multipliers.
\item LMP+U coincides with LMP if, for every dispatched generator, the ramping constraints involving binding interval $t$ and the lower generation bound after reserving downward FRP are nonbinding, and at least one dispatched generator is below its upper generation bound after reserving upward FRP. Then $\Delta\pi_t^*=0$, although transmission congestion may remain.
\end{enumerate} 
With no  MWPs, generator $i$'s binding-interval net revenue under LMP+U is
\begin{equation}
\Pi_{it}^{m,\LMPUm}
=\left(\pi_{mt}^{\LMPUm}-c_{it}^{m,\dagger}\right)
g_{it}^{m,R}+\mathcal R_{it}^m.
\label{eq:lmpu_profit}
\end{equation}

\section{Truthful Bidding Incentives}
\label{sec:incentives}
Energy bid-cost recovery alone does not guarantee truthful bidding. We identify conditions under which local offer inflation increases binding-interval net revenue under LMP+MWP but leaves it constant under LMP+U. For each offer increase, we re-solve the energy--FRP dispatch and apply LMP+U, allowing energy prices and FRP payments to respond.

Fix a dispatched generator $i$ at bus $m$ and a binding interval $t$. We apply the same offer increase $\delta$ throughout the look-ahead window:
\begin{equation}
c_{i\tau}^m(\delta)=c_{i\tau}^{m,\dagger}+\delta,
\qquad \tau\in\mathcal H_t,\quad\delta\geq0.
\label{eq:bid_perturbation}
\end{equation}
All other offers, forecasts, FRP requirements, physical limits, the preceding implemented dispatch, and true production costs remain fixed.

Suppose the same selected LP basis remains optimal for $0\leq\delta<\epsilon$, for some $\epsilon>0$. Since the offer increase changes only objective coefficients, the generation and FRP awards associated with this basis remain fixed. The corresponding nodal LMPs and FRP payments are affine functions of $\delta$, so their rates of change are constant over this range. 

Suppose another generator $(j,n)\neq(i,m)$ uniquely sets the adder throughout this range. Since its submitted energy offer price remains fixed, \eqref{eq:adjustment}--\eqref{eq:lmpu} give
\begin{equation}
\pi_{mt}^{\LMPUm}(\delta)
=c_{jt}^{n}
+\pi_{mt}^{\LMP}(\delta)-\pi_{nt}^{\LMP}(\delta).
\label{eq:lmpu_setter_relation}
\end{equation}
Thus, the target's LMP+U price responds only through the LMP difference between its bus and the setter's bus. Define the constant sensitivities $\alpha_\ell:=d\pi_{\ell t}^{\LMP}/d\delta$ and $\rho_{it}^m:=d\mathcal R_{it}^m/d\delta$ over this range.

\begin{proposition}[Net-revenue responses to an offer increase]
\label{prop:local_lmpu}
Suppose the same selected LP basis remains optimal and another generator $(j,n)\neq(i,m)$ uniquely sets the adder throughout $0\leq\delta<\epsilon$. Then the target generator's net revenue under LMP+U satisfies
\begin{equation}
\frac{d}{d\delta}\Pi_{it}^{m,\LMPUm}(\delta)
=g_{it}^{m,R}(\alpha_m-\alpha_n)+\rho_{it}^m.
\label{eq:lmpu_profit_change}
\end{equation}
If the target generator receives a positive  MWP under LMP+MWP throughout this range, its net revenue under that settlement satisfies
\begin{equation}
\frac{d}{d\delta}\Pi_{it}^{m,\LMPMWP}(\delta)
=g_{it}^{m,R}+\rho_{it}^m.
\label{eq:mwp_profit_change}
\end{equation}
\end{proposition}

With a positive MWP, energy revenue plus the MWP follows the submitted offer; under LMP+U, energy revenue follows the LMP difference between the target's and setter's buses. If this difference and the target's FRP payment remain constant, offer inflation increases LMP+MWP net revenue but leaves LMP+U net revenue constant within the stated range. Generators at the same bus automatically have equal LMP responses.  This result is illustrated in Sec.~\ref{subsec:sim_incentives}.



\begin{figure}[!t]
\centering
\begin{minipage}[c]{0.50\columnwidth}
    \centering
    \includegraphics[width=\linewidth]{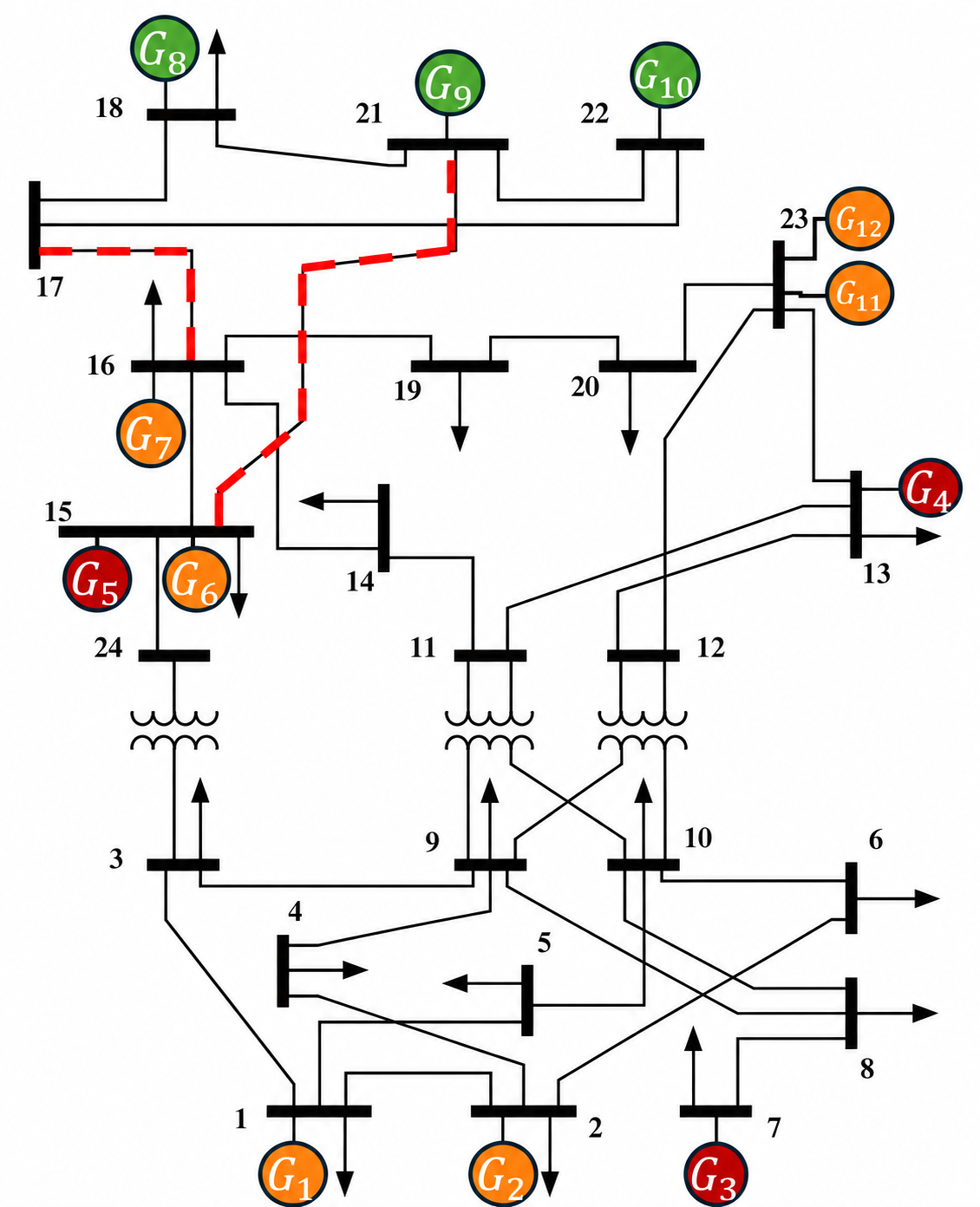}
\end{minipage}\hfill
\begin{minipage}[c]{0.4\columnwidth}
    \centering
    \scriptsize
    \setlength{\tabcolsep}{2pt}
    \renewcommand{\arraystretch}{0.93}
    \begin{tabular}{@{}crc@{}}
        \toprule
        Generator & Cost & Group \\
                  & (\$/MWh) & \\
        \midrule
        $G_1$    & 13.32 & Medium \\
        $G_2$    & 13.32 & Medium \\
        $G_3$    & 20.70 & High \\
        $G_4$    & 20.93 & High \\
        $G_5$    & 26.11 & High \\
        $G_6$    & 10.52 & Medium \\
        $G_7$    & 10.52 & Medium \\
        $G_8$    &  6.02 & Low \\
        $G_9$    &  5.47 & Low \\
        $G_{10}$ &  0.00 & Low \\
        $G_{11}$ & 10.52 & Medium \\
        $G_{12}$ & 10.89 & Medium \\
        \bottomrule
    \end{tabular}
\end{minipage}
        \vspace{-0.3em} 
\caption{IEEE RTS-24 network, marginal production costs, and generator groups. Green, orange, and red indicate low-, medium-, and high-cost generators; dashed red lines indicate the lines with reduced limits in the congested case.}
\label{fig:rts24_network}
\end{figure}

 \begin{figure}[htbp]
    \centering
    \begin{minipage}{\columnwidth}
        \centering
        \includegraphics[width=\textwidth]{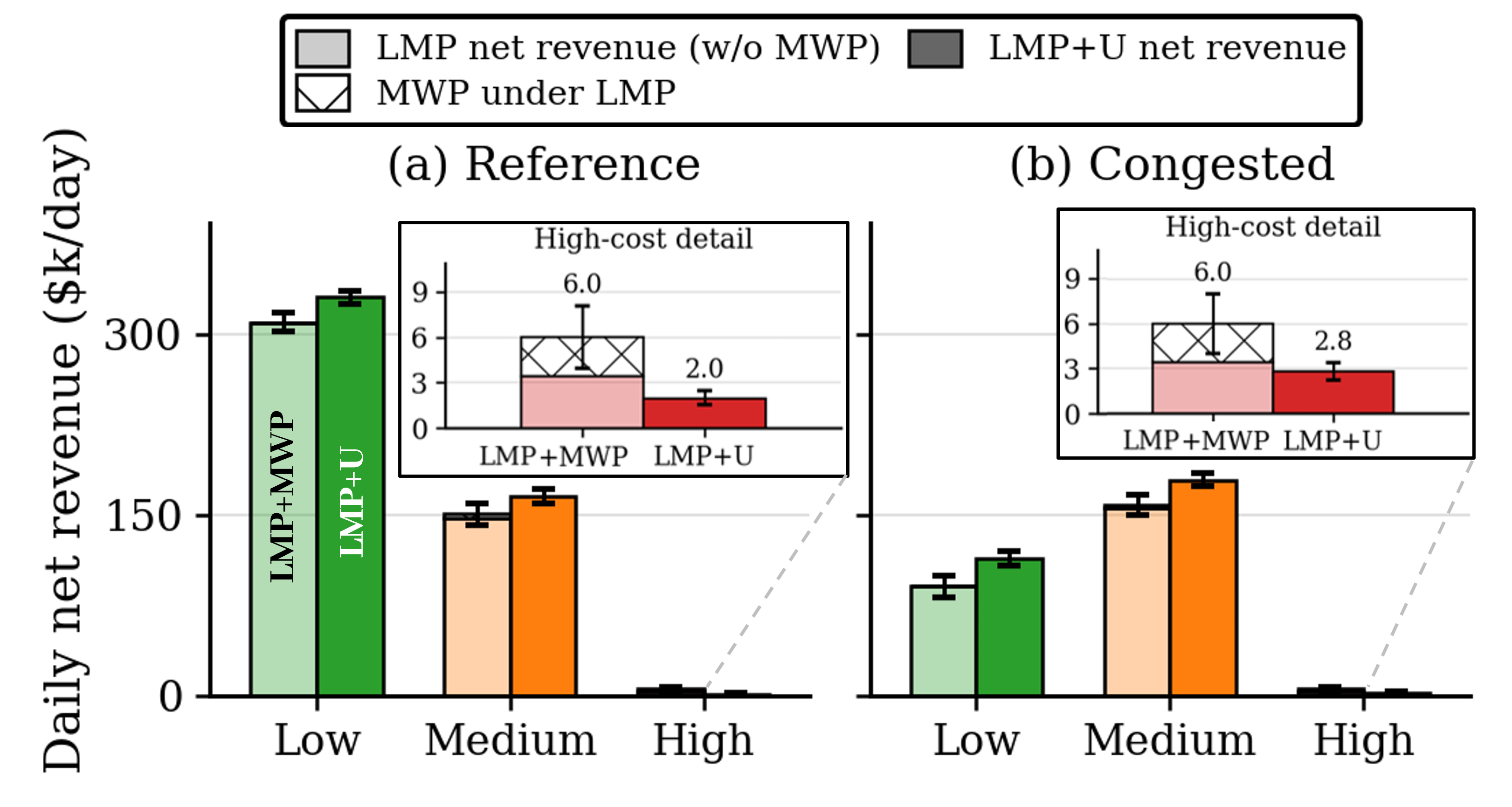}
        
        \vspace{-1.5em} 
        
        \caption{Mean daily group net revenue with FRP and original ramping limits in (a) the reference and (b) congested cases. Patterned portions represent MWPs. Error bars show 95\% confidence intervals for means across 100 trajectories.}
        \label{fig:fig2}
        
        \vspace{-0.5em} 
    \end{minipage}
\end{figure}

\section{Numerical results}

\subsection{Simulation Settings}
\label{sec:simulation_settings}
We simulated the 12-generator IEEE RTS-24 system (Fig.~\ref{fig:rts24_network}) using generator data and nodal load shares from \cite{Ordoudis2016RTS24}, with minimum generation limits set to zero. Dispatch used a four-interval rolling window ($W=4$) with 15-minute intervals over one day.

We generated 100 net-demand trajectories using the Gaussian AR(16) residual model in \cite{Chen26arxiv}, fitted to historical CAISO net demand. The generated deviations were scaled and added to the RTS-24 daily load profile. Binding-interval demand was known, while advisory forecasts were updated recursively. FRP requirements combined forecast demand changes with nominal 95\% Gaussian uncertainty margins for the next interval. We also considered a case with both FRP requirements set to zero (FRP0), using the same demand trajectories and physical limits.

Original ramping limits were 10\% of generator capacity per interval. We separately increased each group's aggregate capability by 30, 75, or 150~MW per interval, allocating ramping capability in proportion to generator capacities.

The \emph{reference case} used the original transmission limits \cite{Ordoudis2016RTS24}; the \emph{congested case} reduced limits on lines 15--21 and 16--17 from 1,000 and 500~MW to 450 and 400~MW, respectively.

For each dispatch, LMP+MWP and LMP+U used the same generation and FRP payments. We report mean daily net revenues  by group across the 100 trajectories. Total consumer payments include energy payments by demand, energy-only MWPs, and FRP payments. 

\subsection{Effect of FRP Procurement}
With original ramping and reference transmission limits, including FRP increased the fraction of positive adder intervals from 30.82\% to 33.34\% relative to FRP0. Then, FRP increased the frequency of intervals in which at least one dispatched generator's LMP fell below its submitted energy offer.

Relative to FRP0, including FRP reduced the low and medium-cost groups' mean daily net revenues by \$3.2k and \$2.6k under LMP+MWP, respectively, and by \$1.7k and \$1.1k under LMP+U. The high-cost group's net revenue increased under both settlements. Additional FRP procurement therefore did not translate into higher net revenue for every group.

With FRP, LMP+U yielded higher net revenues than LMP+MWP for the low and medium-cost groups but lower net revenue for the high-cost group (Fig.~\ref{fig:fig2}(a)). For the latter, losses relative to LMP+MWP during negative-adder intervals outweighed gains from periods with higher prices. In general, periods with negative adders reduce energy payments while preserving energy bid-cost recovery. Total consumer payments were \$35.7k/day higher under LMP+U.

\subsection{Effect of Transmission Congestion}
\label{sec:congestion_results}

To examine the effect of transmission congestion, we compared the reference and congested cases with FRP and original ramping limits. Tightening transmission limits reduced the low-cost group's mean daily net revenue from \$330.6k to \$113.8k under LMP+U and from \$309.9k to \$91.0k under LMP+MWP (Fig.~\ref{fig:fig2}). Energy bid-cost recovery therefore did not protect these generators' net revenues from congestion.

In the congested case, $G_5$ uniquely set every negative adder, lowering energy payments to all dispatched generators while preserving their energy bid-cost recovery. Despite these reductions, total consumer payments under LMP+U averaged \$43.4k/day more than under LMP+MWP.

\subsection{Effects of Increasing Ramping Capability}
\label{sec:flexibility_results}

 \begin{figure}[htbp]
    \centering
    \begin{minipage}{\columnwidth}
        \centering
        \includegraphics[width=\textwidth]{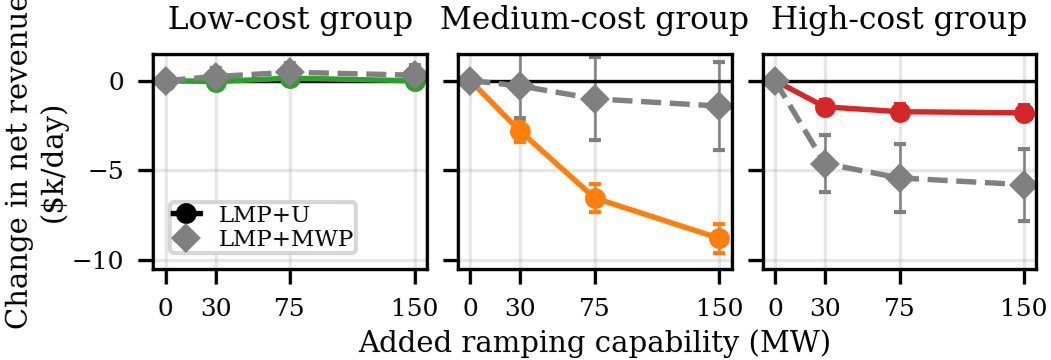}
        
        \vspace{-1.0em} 
        
        \caption{Mean daily net-revenue changes from separately increasing each group's ramping capability, with FRP and reference transmission limits. Error bars show 95\% confidence intervals for mean changes across 100 paired trajectories.}
        \label{fig:group_flexibility}
        
        \vspace{-0.1em} 
    \end{minipage}
\end{figure}

With FRP and reference transmission limits, increasing ramping capability did not necessarily increase the group's net revenue (Fig.~\ref{fig:group_flexibility}). Adding 150~MW per interval in each direction to the medium-cost group reduced its mean daily net revenue by \$8.80k under LMP+U and \$1.42k under LMP+MWP. Under LMP+U, lower adder payments outweighed gains in combined energy and FRP revenue after production costs.

For the high-cost group, the same increase reduced mean daily net revenue by \$1.80k under LMP+U and \$5.81k under LMP+MWP. Higher adder payments partly offset reductions in energy net revenue under LMP+U. The low-cost group's net revenue changed little. These group-level results suggest ramping withholding incentives but do not establish profitable withholding by individual generators.

\subsection{Illustration of Truthful Bidding Incentives}
\label{subsec:sim_incentives}
To illustrate Proposition~\ref{prop:local_lmpu}, we used scenario~45, interval~38, with high-cost ramping limits at 3\% of capacity per interval and line limits at 65\% of their reference values. Under truthful bidding, $G_4$ at bus~13 produced 352.09~MW at a marginal cost of \$20.93/MWh and an LMP of \$10.52/MWh. Generator $G_5$ at bus~15 uniquely set the adder, giving $G_4$ an LMP+U price of \$26.11/MWh.

With other inputs fixed, raising $G_4$'s offer throughout the look-ahead window retained the selected optimal basis, $G_5$'s unique setter status, and a positive energy-only MWP for $0\leq\delta<5.18$~\$/MWh. Since $\alpha_{13}=\alpha_{15}$ and $\rho_{4t}^{13}=0$, LMP+U net revenue remained constant, while each \$1/MWh offer increase raised LMP+MWP net revenue by \$88.02 over the 15-minute binding interval.

\section{Conclusion}
\label{sec:conclusion}

We extended LMP+U to transmission-constrained energy and FRP dispatch, eliminating energy-only MWPs at minimum demand energy payments while preserving nodal LMP differences and energy merchandising surplus, with FRP payments settled separately. Simulations showed that additional ramping capability and congestion can reduce generator net revenues, while LMP+U increases consumer payments relative to LMP+MWP. Under identified local conditions, offer inflation increases LMP+MWP net revenue but leaves LMP+U net revenue constant.
\vspace{-0.2em}
\section*{Acknowledgment}
The authors used OpenAI's ChatGPT for feedback on readability and edited Sections I--VI accordingly. The arguments and results are original and the authors' own.


\end{document}